\documentstyle[epsfig,subfigure]{europhys}
\def\And{{\rm and\ }}

\newif\ifboo \boofalse

\def\Review#1{\boofalse{\it #1},}
\def\Name#1{{\sc #1},}
\def\Vol#1{\ifboo Vol. {\bf #1}\else{\bf #1}\fi}
\def\Year#1{\ifboo #1\else(#1)\fi}
\def\Book#1{\bootrue{\it #1},}
\def\Page#1{\ifboo {\rm p. #1}\else{\rm #1}\fi}
\newcommand\bm[1]{\mbox{\boldmath $#1$\unboldmath}} 
\euro{49}{2}{224-230}{2000}
\Date{November 24, 1999}
\shorttitle{E. MARIANI {\it et al.}: NEW SELECTION RULES FOR RESONANT RAMAN...}
\title{New Selection Rules for Resonant Raman Scattering on Quantum Wires.}
\author{E. Mariani
\footnote{\noindent present address: I. Institut f\"ur Theoretische
    Physik, Universit{\"a}t Hamburg\\
    Jungiusstra{\ss}e 9, D-20355 Hamburg, Germany.}\inst{1},
  M. Sassetti\inst{1},
  and B. Kramer\inst{2}}
\institute{\inst{1}   Dipartimento di Fisica, INFM,
  Universit{\`a} di Genova\\
  Via Dodecaneso 33, I-16146 Genova, Italy\\
\inst{2} I. Institut f\"ur Theoretische Physik,
  Universit{\"a}t Hamburg\\
  Jungiusstra{\ss}e 9, D-20355 Hamburg, Germany}

\rec{dd July 1999}{dd Month 1999 in final form}


\begin{document}
\maketitle

\begin{abstract} The bosonisation technique is used to calculate the resonant 
  Raman spectrum of a quantum wire with two electronic sub-bands
  occupied.  Close to resonance, the cross section at frequencies in
  the region of the inter sub-band transitions shows distinct peaks in
  {\em parallel polarisation} of the incident and scattered light
  that are signature of collective higher order {\em spin density
    excitations}. This is in striking contrast to the conventional
  selection rule for non-resonant Raman scattering according to which
  spin modes can appear only in perpendicular polarisation. We predict
  a new selection rule for the excitations observed near resonance,
  namely that, apart from charge density excitations, only spin modes
  with positive group velocities can appear as peaks in the spectra in
  parallel configuration close to resonance. The results are
  consistent with all of the presently available experimental data.
\end{abstract}

Raman spectroscopy is one of the most powerful experimental techniques
for investigating the dispersion relation of elementary excitations in
condensed matter \cite{pa88,k75}. Recently, the spectra of the
strongly correlated electrons in quantum wires
\cite{gwpetal91,sgpetal94,sbketal96} have been the subject of
considerable experimental effort. It has been
suggested that intraband charge %\cite{sh94} 
and spin density excitations (CDE and SDE) which are signature of
Luttinger liquid behaviour \cite{s93} can be identified in parallel
and perpendicular polarisations of incoming and scattered light,
respectively. In addition, the peaks in the polarised {\em resonant}
Raman spectra, when the energy of the incident photon approaches the
energy gap, that were associated so far with ``single particle
excitations'' (``SPE'') have been identified as collective, higher
order spin density excitations in the region of the {\em intraband}
modes \cite{sk98}. In order to treat the interband spectra, the
bosonisation approach has been generalised to include interband
coupling terms \cite{snk99}. It has been shown that the essential
experimental features observed {\em far from resonance} in the
polarised and depolarised configurations of the light beams, in the
frequency region of the lowest {\em interband} CDE and SDE, can be
recovered.

In the present paper, we show that in the region of the interband
transitions, higher-order SDE give rise to pronounced peaks in the
resonant Raman spectra in {\em parallel polarisation}, in analogy with
those found for the intraband modes (so-called ``SPE''). We predict
that these peaks are subject to a new selection rule: only interband
SDE with a positive group velocity can appear in the resonant Raman
spectra in the {\em parallel configuration} as sharp peaks.  All of
the presently available experimental data of quantum wires are
completely consistent with our predictions.  Nevertheless, more
detailed experimental investigations, especially on quantum wires with
two sub-bands, are necessary in order to confirm our results
quantitatively. This would provide further strong evidence for the
importance of the Coulomb interaction and the collective and bosonic
nature of the electronic excitations in quantum wires which are
signature of Luttinger liquid behavior \cite{sk98a}.

We consider a quantum wire of length $L$ (periodic boundary
conditions) with two occupied conduction sub-bands with dispersion
relations $ E_{i}(k)=\varepsilon_{i}+\hbar^{2}k^{2}/2m $ ($i=1,2$)
where $\varepsilon_{1}=0$, $\varepsilon_{2}=E_{0}$, $E_{0}$ the
intersub-band separation, and $m$ the effective mass of an electron in
the conduction band. The position of the Fermi level $E_{\rm F}$
defines the Fermi velocities $v_{\rm F1,2}$ corresponding to the two
sub-bands. We assume a three dimensional (3D) Coulomb interaction.
When projected with the confining wave functions onto the sub-bands,
this gives the effective 1D interaction between the electrons
\cite{sk98}. The Hamiltonian of the free electrons in terms of Fermion
operators $c^{\dagger}_{is}(k)$, $c_{is}(k)$ (spin $s$ and wave number
$k$) is
\begin{equation}
\label{H0}
H_{0}=\sum_{i=1}^{2}\sum_{s,k}E_{i}(k)c^{\dagger}_{is}(k)
c_{is}(k)  
\end{equation}
and the interaction energy
\begin{equation}
\label{Hint}
H_{\rm int}=\sum_{ijlm}\sum_{ss^{\prime}}\sum_{qkk^{\prime}}
V_{ijlm}(q)\,c^{\dagger}_{is}(k+q)c^{\dagger}_{js^{\prime}}
(k^{\prime})c_{ls^{\prime}}(k^{\prime}+q)c_{ms}(k)
\end{equation}
where $V_{ijlm}(q)$ are the matrix elements of the interaction potential.

In order to diagonalise this Hamiltonian, we use a generalisation of
the Luttinger model which is commonly employed to describe the
small-momentum collective excitations of 1D systems of interacting
electrons \cite{l63}. The presence of two sub-bands allows pair
excitations both within a given sub-band ({\em intraband}
excitations) and between different sub-bands ({\em inter-band}
excitations). As for the Luttinger model, we linearise the dispersion
relations near the Fermi level. One obtains two spectral    branches 
$\lambda=\pm$ for right/left-moving electrons. The
Hamiltonian can then be diagonalised in terms of the density operators
$\rho_{ij,s}^{\lambda}(q)=\sum_{k}
c^{\lambda\dagger}_{is}(k+q)c^{\lambda}_{js}(k)$.

\begin{figure}[h]
%\vspace{7cm}
\begin{picture}(400,210)
\put(5,0){\epsfxsize=14cm\epsffile{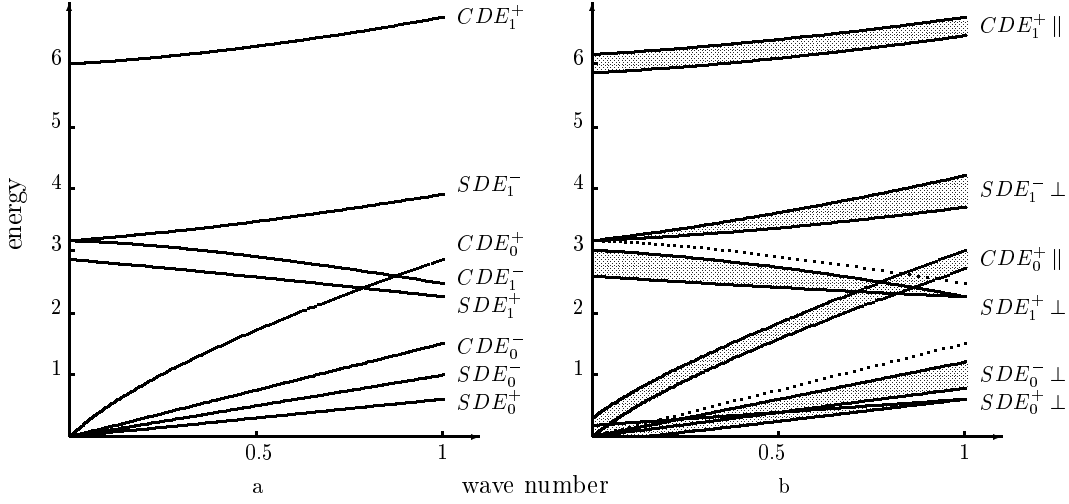}}
%\put(40,90){\framebox(10,5){Figure1.ps}}
%\includegraphics[bb=-20 -20 300 100,clip]{Figura1.ps}
%\put(83,60){$E_{F}$}
%\put(302,66){$\omega$}
%\put(302,122){$i=1$}
%\put(302,147){$i=2$}
%\put(240,-7){$q$}
%\put(144,18){$E_{0}$}
\end{picture}
\caption[1]{(a) Dispersion relations of intra- and
  inter-band CDE and SDE of the two-band Luttinger model. Energy in
  meV; wave number in 10$^{5}$cm$^{-1}$.  We chose $\hbar v_{\rm
    F1}=$10$^{-5}$ meV$\cdot$cm and $v_{\rm F2}=0.6\:v_{\rm F1}$.  (b)
  Frequency-wave number dependence of the peaks in the non-resonant
  Raman spectra for the modes shown in (a). In parallel polarisation
  of incoming and scattered light, only intra- and interband charge
  density modes, CDE$_0^{\pm}$ and CDE$_1^{\pm}$ ($\pm$ symmetric and
  anti-symmetric excitations) can be observed. In perpendicular
  polarisation, SDE$_0^{\pm}$ and SDE$_1^{\pm}$ appear.  Widths of the
  shaded areas: approximate intensity of the peaks in the Raman cross
  section. Dotted branches: intensity of Raman peaks unobservably
  small.}
\label{fig:1}
\end{figure}

By using a mean field assumption the Hamiltonian can be decomposed
into {\em independent} intraband and inter-band contributions
\cite{snk99} with independent charge and spin densities
\begin{equation}
  \label{eq:3}
  \rho_{ij}=\rho_{ij,\uparrow} + \rho_{ij,\downarrow}\,,\qquad
  \sigma_{ij}=\rho_{ij,\uparrow} - \rho_{ij,\downarrow}
\end{equation}
respectively, similar to the one-band Luttinger model. The Coulomb
repulsion affects the CDE while the SDE feel only the exchange
interaction. The system can be diagonalised by using a generalised
Bogolubov transformation \cite{ps93}. For parameters characteristic
for state-of-the-art semiconductor quantum wires
\cite{gwpetal91,setal99}, the dispersion relations are shown in
Fig.~\ref{fig:1}a.

The four energetically lowest branches of the spectrum correspond to
the intraband charge and spin density modes, CDE$_0^{+,-}$ and
SDE$_{0}^{+,-}$, respectively. The two lowest branches are the SDE.
They are linear in the wave number with slopes approximately equal to
the Fermi velocities $v_{\rm F1,2}$ since the exchange interaction is
very small. The next-higher two branches, starting again at 
$\omega = 0$ when $q = 0$, are the charge density modes. The uppermost 
intraband
CDE, symmetric in the charge densities of the two sub-bands,
reflects the Fourier-transform of the interaction, $\omega(q)\approx
\left|q\right|(v_{{\rm F}1}+v_{{\rm F}2})\left[1+2V_{1111}(q)/\pi\hbar(v_{{\rm
      F}1}+v_{{\rm F}2})\right]^{1/2}$ in the limit of long wave length 
\cite{s93,sk98}. 
The slope of the energetically lower (anti-symmetric) CDE is proportional the
geometrical average of the two Fermi velocities, $\omega(q)\propto
(v_{{\rm F}1}v_{{\rm F}2})^{1/2}\left|q\right|$ \cite{s93,sk98}.

The four modes which start at $q=0$ with finite frequencies are
interband charge and spin excitations. The highest branch with the
positive group velocity ${\rm d}\omega(q)/{\rm d}q$ corresponds
to a symmetric interband charge density mode $\approx
\rho_{12}^{+}+\rho_{12}^{-}$. It is strongly shifted to higher energy
by the Coulomb repulsion (depolarisation shift). The anti-symmetric
branch, $\approx \rho_{12}^{+}-\rho_{12}^{-}$, does not feel the
Coulomb repulsion at small wave numbers and has a negative group velocity. 
It is degenerate with the anti-symmetric interband SDE, 
$\approx  \sigma_{12}^{+}-\sigma_{12}^{-}$ at $q=0$, 
at an energy which is given
by the interband spacing $E_{0}$ plus the exchange energy
$V_{\rm ex}$. In our example (Fig.\ref{fig:1}) $E_{0}=3$ meV and 
$V_{\rm ex}\approx 0.2$ meV.  
The two remaining branches of the spectrum
correspond to the symmetric and anti-symmetric SDE ($\approx
\sigma_{12}^{+}\pm\sigma_{12}^{-}$), starting at $\hbar \omega \approx
E_{0} \mp V_{\rm ex}$, respectively. 
%As the exchange interaction is negative, the anti-symmetric SDE is higher 
%in frequency.

The differential cross section for the inelastic scattering of the
light is given by Fermi's golden rule and can be written in terms of a
response function $\chi(q,\omega)$ \cite{k75},
\begin{equation}
\label{Sigmachi} \frac{{\rm d}^{2}\sigma}{{\rm d}\Omega
    {\rm d}\omega}=\left(\frac{e^{2}}{mc^{2}}\right)^{2}
  \frac{\omega_{\rm O}}{\omega_{\rm I}}
  \:\frac{n(\omega)+1}{\pi}\:{\rm Im}\chi
  (q,\omega)               \end{equation}
with $\omega =\omega_{\rm I} -\omega_{\rm O}$ the difference between
the frequencies of the incident and scattered photons, $q=k_{\rm
  I}-k_{\rm O}$ the transferred wave number, $n(\omega)$ the Bose
distribution, and 
\begin{equation}
\label{chi}
 \chi(q,t)={\rm i}\,\Theta (t)\langle
 \left[ N^{\dagger}(q,t),N(q,0)\right]\rangle
\end{equation}
where $\langle \ldots \rangle$ is the thermal average.

The operator
\begin{equation}
\label{Nq}
  N(q,t)=\gamma_{0}\sum_{ij,s}\sum_{\lambda,k}\frac{
    \left({\bm{e}}_{\rm I}    \cdot\bm{e}_{\rm O}+
{\rm i}\left|\bm{e}_{\rm I}\times
    \bm{e}_{\rm O}\right|s\right)}{D_{i}(k,q)}
  \:c^{\lambda\dagger}_{is}(k+q,t)c^{\lambda}_{js}(k,t)
\end{equation}
is a generalised density describing an electronic excitation of
momentum $q$ from band $j$ to band $i$ in the branch $\lambda$.  The
vectors $\bm{e}_{\rm I,O}$ indicate the polarisation of the
incoming/outgoing photon, the constant $\gamma_{0}$ contains the
matrix elements for the transitions between the valence and the
conduction band, and the energy denominator
\begin{equation}
\label{Di}
D_{i}(k,q)=E_{i}(k+q)-E_{\rm v}(k+q-k_{\rm I})-\omega_{\rm I}
\end{equation}
where $E_{\rm v}$ is the energy of the valence band that will be
assumed as dispersion-less. We calculate the above generalised density
by using the bosonization method \cite{sk98} and expanding $D^{-1}$ in
powers of $E_{\rm F}/(E_{\rm g}+\varepsilon_{i}-\hbar\omega_{\rm I} +
\lambda\hbar v_{\rm F12} q/2)$ ($E_{\rm g}=E_{\rm
  g}^{0}+\hbar^{2}k_{\rm F12}^{2}/2m$).  Here, we have defined an
average Fermi wave number, $k_{\rm F12}=(k_{\rm F1}+k_{\rm F2})/2$,
with the corresponding average Fermi velocity $v_{\rm F12}=\hbar
k_{\rm F12}/m$.

If the energy of the incident photon is sufficiently high, we are far
from resonance and can consider only the lowest order term in the
expansion of $D^{-1}$. This gives
\begin{equation}
\label{Noffres}
N(q,t)=\sum_{ij,\lambda}\frac{\gamma_{0}}{E_{\rm g}+\varepsilon_{i}-
               \hbar\omega_{\rm I}+\lambda \hbar v_{\rm F12}q/2}
             \left[\bm{e}_{\rm I}\cdot
               \bm{e}_{\rm O}\:\rho_{ij}^{\lambda}(q,t)+
               i\left|\bm{e}_{\rm I}
                 \times\bm{e}_{\rm O}\right|
               \sigma_{ij}^{\lambda}(q,t)\right]\,.
\end{equation}           
This implies the ``classical selection rule'' for Raman scattering:
only CDE can appear in the polarised configuration and SDE can only be
present in the depolarised one. The behaviour of the peaks in the
frequency-wave number plane to be expected for the non-resonant Raman
spectra, $\left|E_{\rm g}-\hbar\omega_{\rm I}\right|\gg
\varepsilon_{i}+\lambda \hbar v_{\rm F12}q/2$ has been discussed
earlier \cite{snk99}. An overview based on the quantitative results
from our model calculation is given in Fig.~\ref{fig:1}b. Remarkably,
the anti-symmetric charge excitations denoted as CDE$_{0}^{-}$ and
CDE$_{1}^{-}$ turn out to have a very small intensity such that they
cannot be observed far from resonance \cite{sk98}.  Similarly, the
symmetric spin excitations SDE$_{0}^{+}$ and SDE$_{1}^{+}$ can at best
be observed at small wave numbers while SDE$_{0}^{-}$ and
SDE$_{1}^{-}$ will be present at larger wave numbers. As the intraband
SDE are close in frequency, the experiment will show one peak which
increases in frequency almost linearly with $q$.

\begin{figure}[htp]
\vspace{7cm}
\begin{picture}(400,0)
\put(5,0){\epsfxsize=14cm\epsffile{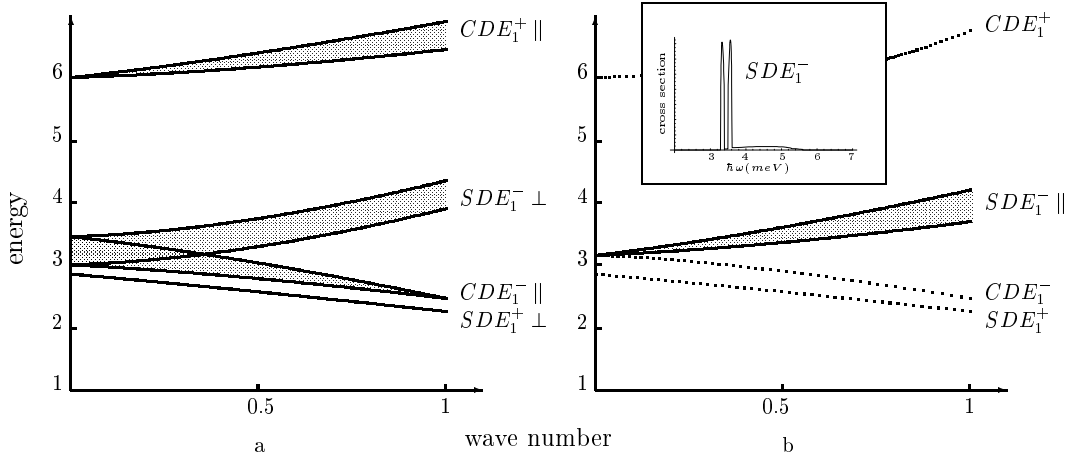}}
\end{picture}
\caption[2]{(a) Behaviour in frequency-wave number
  space of the inter-band resonant Raman peaks according to the
  zero-order term in Eq.~(\ref{Noffres}); energy in meV; wave number
  in 10$^{5}$cm$^{-1}$. (b) Inter-band resonant Raman peaks for
  polarised configuration from the first-order term
  (Eq.~(\ref{DeltaN})); widths of shaded areas: intensities of the
  peaks in the Raman cross section. Dotted branches: intensity of
  Raman peaks unobservably small. Inset: Raman cross section
  (arbitrary units) for the inter-band SDE for $T=0$, with $\hbar
  v_{\rm F1}=$10$^{-5}$ meV$\cdot$cm, $v_{\rm F2}=0.6\:v_{\rm F1}$,
  evaluated at $q=0.8$ 10$^{5}$ cm$^{-1}$. Notice the pronounced
  double peak structure which is due to the inter-band mode with
  positive group velocity (SDE$_{1}^{-}$). The flat
  background-structure between $\hbar \omega \approx 3.5$meV and $\hbar
  \omega \approx 5.5$meV is related to SDE$_{1}^{+}$.}
\label{fig:2}
\end{figure}

Although this is only the first term in the expansion, we nevertheless
have to take into account its contribution towards the spectra when we  
consider photon energies closer to resonance. Indeed, when
$\left|E_{\rm g}+\varepsilon_{i}-\hbar\omega_{\rm I}\right| 
\approx \hbar v_{\rm F12}q/2$ the
term can contribute considerably. The behaviour in frequency-wave
number space of the Raman peaks corresponding to the inter-band
transitions according to this zero-order term is shown in
Fig.~\ref{fig:2}a \cite{snk99}. In parallel polarisation, only CDE are
observable: CDE$_{1}^{-}$ is expected to dominate at small $q$, and
CDE$_{1}^{+}$ becomes strong for larger $q$. In perpendicular
configuration, only SDE$_{1}^{-}$ is expected.

Using the bosonic representation of the fermionic operators in Eq. 
(\ref{Nq}), the first-order contribution to $N(q,t)$ is
\newpage
\begin{eqnarray}\label{DeltaN}
\Delta N(q,t)&=& -\frac{\pi}{2L}\frac{\gamma_{0}\hbar^{2}}{m}
               \sum_{ij,\lambda}
               \:\frac{k_{\rm F12}+\lambda q/2}
               {\left[E_{\rm g}+\varepsilon_{i}-\hbar\omega_{\rm I}+
                   \lambda \hbar v_{\rm F12}q/2\right]^{2}}\times
\nonumber \\
               &\times&\sum_{k,l}\left\{(\bm{e}_{\rm I}
                 \cdot\bm{e}_{\rm O})
                 \left[\,\rho^{\lambda}_{ll}(q-k,t)     
                   \rho^{\lambda}_{ij}(k,t)+
                   \sigma^{\lambda}_{ll}(q-k,t)
                   \sigma^{\lambda}_{ij}(k,t)\,\right]+\right.
\nonumber\\
               &+&\left. {\rm i}\:\left| \bm{e}_{\rm I}
                   \times\bm{e}_{\rm O}
                 \right|\left[\,\rho^{\lambda}_{ll}(q-k,t)
                   \sigma^{\lambda}_{ij}(k,t)+
                   \sigma^{\lambda}_{ll}(q-k,t)
                   \rho^{\lambda}_{ij}(k,t)\,\right]\right\}\,.
\end{eqnarray}
The intraband case ($i=j$) has already been studied in detail and can
be treated within the one band Luttinger model \cite{sk98}. It has
been found that the spin density operators present in the polarised
term ($\propto \bm{e}_{\rm I} \cdot\bm{e}_{\rm
  O}$) give rise to Raman peaks that violate the classical
selection rule, since they appear in parallel polarisation. In
previous works, these have been associated with ``single particle  
excitations''.

The present two-band model specifically allows for investigating the
behaviours of the inter-band modes closer to resonance.  The polarised
terms in Eq.~(\ref{DeltaN}) ($\propto \bm{e}_{\rm I}
\cdot\bm{e}_{\rm O}$) are bi-linear in the spin
density operators and yield the dominant contribution to the
Raman cross section. An important feature is that they couple directly
inter- and intraband modes.

By writing them in terms of the eigen-modes of the Hamiltonian, for
which the time evolutions are known, and by evaluating the response
function Eq.~(\ref{chi}) we obtained a closed expression for ${\rm
  d}^{2}\sigma/{\rm d}\Omega {\rm d}\omega$. Figure~\ref{fig:2}b shows
the behaviour in frequency-wave number space of the possible interband
Raman peaks according to the first-order term, Eq.~(\ref{DeltaN}).
Only the anti-symmetric spin mode SDE$_{1}^{-}$ contributes in
parallel polarisation! Due to the mixing of this inter-band mode which   
has a positive group velocity with the two intraband SDE related to
the two Fermi velocities a double peak appears in the spectrum at zero
temperature (Fig.~\ref{fig:2}b inset). The width of the peaks is
roughly proportional to the difference between the group velocities of
the coupled modes and they are centred on two lines with slopes equal
to the two Fermi velocities $v_{\rm F1}$ and $v_{\rm F2}$, starting around
$E_{0}+V_{\rm ex}$.  For non-zero temperatures, the two peaks are
smeared and appear as only one structure. The presence of disorder will
also broaden the peaks. The resulting SDE-peak in the parallel
configuration appears roughly at the frequency of the SDE$_{1}^{-}$,
and its intensity increases with increasing wave number as indicated
by the increasing width of the shaded area in Fig.~\ref{fig:2}b.

On the other hand, by coupling the inter-band spin mode with negative group
velocity to the two intraband spin density excitations we obtain a
very flat and broad structure (Fig.~\ref{fig:2}b inset) that will not
be observable experimentally.
                                
In the terms in Eq.~(\ref{DeltaN}) which correspond to perpendicular
polarisation ($\propto \bm{e}_{\rm I} \times
\bm{e}_{\rm O}$) charge and spin density operators are
mixed. These contribute towards the Raman cross section only with
broad structures within frequency windows that are given by the
differences of the group velocities of the modes that are mixed.
They are not expected to be experimentally detectable.

In summary, we have evaluated the Raman cross section of a
semiconductor quantum wire with two conduction sub-bands
occupied by using the bosonisation method for two coupled Luttinger
liquids. We have found strong violations of the ``classical selection
rules'' for Raman scattering. Due to higher order terms which become
important when the energy of the incident photons approaches the
energy gap between conduction and valence band of the semiconductor,
SDE modes appear in perpendicular as well as in
parallel polarisation of incident and scattered photons. 
For these, a novel selection rule has been established: only SDE which 
can combine
with intraband modes with similar group velocities can occur in
parallel polarisation as distinct peaks.  On the other hand, in
perpendicular configuration, no higher-order CDE modes become
observable.

Our model allows to understand the physical reason for this selection
rule: coupling between an inter-band SDE with positive group velocity
and intraband modes which have group velocities ($\approx$ Fermi
velocity) leads to a long-range correlation in time between the
corresponding collective excitations. As a consequence, $\chi
(q,\omega)$, the Fourier transformed correlation function, will show a
pronounced structure in ($q,\omega$)-space. On the other hand, due to
opposite signs of the group velocities of the inter-band mode
SDE$_{1}^{+}$ and of the two intraband excitations, the correlation in
time is short-range. Its Fourier transform will be long-range in
frequency. A similar qualitative argument explains also the {\em
  absence} of pronounced, sharp peaks in perpendicular polarisation
due to the combined spin-charge excitations which have necessarily
very different group velocities. As the origin of structure in
perpendicular polarisation is related to spin-orbit coupling, no
higher-order CDE alone can occur. Only higher-order CDE accompanied by
at least one SDE are present. This physical argument suggests that the
new selection rule is very probably also valid for more than two
sub-bands.

Our findings are consistent with existing Raman data of two-band
quantum wires \cite{gwpetal91}, where the structures violating the
selection rules have been denoted as ``single particle excitations''.
However, more precise measurements are needed in order to verify our
quantitative predictions concerning, for instance, the dependence of
the peak intensities on wave number and frequency (Fig.~\ref{fig:2}).

In conclusion, previous and present experimental and theoretical
findings indicate that the observed features in the Raman spectra of
quantum wires seem to be completely understandable in terms of the
collective intra- and inter-band spin and charge excitations
characteristic of coupled Luttinger liquids. This represent in our
opinion strong evidence that the Luttinger model in connection with
bosonisation correctly describes the collective modes in quantum wires
for $q\neq 0$. Whether or not this is also valid for $q \to 0$ (near
$T=0$) is an open question and presently intensively discussed
\cite{fl93,f93,s96}.

{\em Acknowledgements:} We thank Franco Napoli for useful discussions.  
The work has been supported by the Italian MURST Cofin 98, the Deutsche
Forschungsgemeinschaft via SFB 508 and Graduiertenkolleg
``Nanostrukturierte Festk\"orper'' of the Universit\"at Hamburg, and by
the EU via the TMR programme.
\\
\\

\end{document}